\documentclass[aps,prd,showpacs,notitlepage,nofootinbib,superscriptaddress,floatfix,showkeys,twocolumn]{revtex4-1}
\usepackage{float}
\usepackage{multirow}
\usepackage{siunitx}
\usepackage{verbatim}
\usepackage{graphicx}
\usepackage{amsmath}
\usepackage[usenames]{color}
\usepackage{amssymb}
\usepackage{hyperref}
\newcommand{\be}{\begin{equation}}
\newcommand{\ee}{\end{equation}}
\newcommand{\bea}{\begin{aligned}}
\newcommand{\eea}{\end{aligned}}
\newcommand{\pr}{\partial}

\newcommand{\bse}{\begin{subequations}}
\newcommand{\ese}{\end{subequations}}

\newcommand{\bmm}{\begin{multline}}
\newcommand{\emm}{\end{multline}}

\newcommand{\mi}{\mathrm{i}}
\begin{document}
\title{Superradiant scattering of massive scalar field due to magnetically charged rotating black hole}
\author{Rajesh Karmakar}
\email{rajesh018@iitg.ac.in}
\affiliation{Department of Physics, 
Indian Institute of Technology Guwahati, Assam 781039, India}

\begin{abstract} 
Rotating black holes are well known to amplify the perturbing bosonic fields in certain parameter spaces. This phenomenon is popularly known as superradiance. In addition to rotation in the spacetime, charge plays a crucial role in the amplification process. In the present study, we have considered the spacetime of a magnetically charged rotating black hole emerging from the coupling of nonlinear electromagnetic field configuration to gravity. Due to this black hole spacetime, we have studied the scattering states of a massive scalar field and investigated the superradiant amplification process. We find that the magnetic charge of the spacetime affects the magnitude of the amplification and significantly enlarges the allowed frequency ranges for which superradiance happens. In comparison, our analysis reveals that some of the behaviours are quite similar to what has been found in the case of the Kerr-Newmann black hole. Moreover, we observe exact similarities in specific parameter spaces, which could imply a probable correspondence between the magnetically charged black hole and the Kerr-Newman black hole in certain scenarios. Such similarities could have important implications for observing the effects of the non-linear nature of electromagnetic fields in the vicinity of spinning black holes. In addition, we have investigated the superradiant instability regime due to the massive potential barrier and compared our results with that of the Kerr-Newmann black hole.
 
\end{abstract}

\maketitle

\newpage

\section{Introduction}\label{intro}

Since Schwarzschild discovered the first exact solution \cite{Schwarzschild:1916uq} of Einstein equation in the general theory of relativity (GR), the study of black holes (BHs) has advanced significantly in both theoretical and observational aspects. Now Schwarzschild spacetime is known to be the only spherically symmetric, static vacuum solution according to Birkhoff's theorem \cite{Carroll:2004st}. Following the discovery of Schwarzschild spacetime, Kerr introduced a rotating vacuum solution \cite{Kerr:1963ud}, which is considered well-suited to describe astrophysical BHs, formed as the final state of gravitational collapse. In the stellar mass range, phenomenological studies of Kerr BH have been remarkably consistent with the observation of gravitational waves (GWs) by LIGO-Virgo-KAGRA (LVK) \cite{LIGOScientific:2016aoc, LIGOScientific:2016emj, LIGOScientific:2016vbw, LIGOScientific:2017ycc, LIGOScientific:2017vwq, LIGOScientific:2020stg, LIGOScientific:2020aai, LIGOScientific:2020zkf, LIGOScientific:2023fpk, KAGRA:2023pio, KAGRA:2022twx} over the past few years. On top of this, the Event-Horizon-Telescope (EHT) collaboration's observations of the shadow images of Sgr A* and M87* \cite{EventHorizonTelescope:2019dse, EventHorizonTelescope:2022wkp} have further validated the Kerr solution's effectiveness in describing supermassive BHs. In addition to rotation in spacetime, the electric charge parameter was later incorporated, leading to what is known as the Kerr-Newman (KN) BH spacetime \cite{Newman:1965my}. Constraints on the charge-to-mass ratio of this BH have been analyzed through GW signal \cite{Carullo:2021oxn} and shadow image observations \cite{Ghosh:2022kit}, revealing that the ratio is very small. Nevertheless, following the no-hair theorem, KN BH represents the most general stationary vacuum solution derived from the Einstein-Maxwell equation \cite{HeuslerM}. However, all these vacuum solutions are endowed with unphysical curvature singularity, whose occurrence is inevitable according to the cosmic censorship conjecture proposed by Hawking and Penrose \cite{Hawking:1973, Penrose:1969pc, Hawking:1979ig, Wald:1997wa}. People have resorted to investigating this issue in larger scenarios by incorporating modifications to Einstein's gravity theory.

In the context of modified gravity theories, charged BHs sourced by nonlinear electrodynamics (NED) have recently gained more significance given the advancement in astrophysical observations. This is due to the fact that astrophysical BHs are often surrounded by strong magnetic fields or immersed in a plasma medium. Largely, the NED-sourced BHs can be categorised into two classes: regular BH (RBH) and singular BH (which we refer to as simply BH). Originally initiated by Bardeen \cite{Bardeen1968}, NED-sourced RBHs, like other RBHs, are free from curvature singularity, however, possess a coordinate singularity at the horizon in the usual manner (see \cite{Bambi:2023try} for further details). There have been various proposals over the years for RBHs since then \cite{Hayward:2005gi, Ayon-Beato:1998hmi, Fan:2016hvf} and have drawn much attention in recent times \cite{Riaz:2022rlx, Villani:2021lmo, Allahyari:2019jqz, Abdujabbarov:2016hnw}. However, their viability and stability remain subject to much debate \cite{Carballo-Rubio:2018pmi, Franzin:2022wai}. In the singular BH category, Einstein-Born-Infeld (EBI) \cite{Born:1934gh} and Einstein-Euler-Heisenberg (EEH) \cite{Yajima:2000kw, Breton:2019arv} spacetime have been actively studied over the years \cite{Breton:2003tk, Breton:2016qyf, Allahyari:2019jqz}, among others. Recently, Ghosh and Kumar \cite{Ghosh:2021clx} proposed a magnetically charged rotating BH (which we refer to as GK BH) utilizing the NED coupling to gravity. This BH has the same type of singularities as exhibited by Kerr BH. The shadow cast by this BH has been analyzed in \cite{Vagnozzi:2022moj}. Apart from the observational significance, this BH spacetime is closer to KN BH as compared to newly proposed, NED-motivated rotating RBHs \cite{Franzin:2022wai, Bambi:2013ufa}, therefore, could be a better playground for investigating the electromagnetic duality in the semiclassical process \cite{Deser:1996xp, Hawking:1995ap, Liu:2019smx}. This has particularly motivated us to study this GK BH spacetime. 

In the present work, we will analyse the behaviour of the massive scalar perturbation to the magnetically charged GK BH and investigate the implication of this charge on the superradiant amplification process.  For such charged BHs, it is more interesting to consider a charged scalar field \cite{Semiz:2005gs}. However, to simplify the discussion, we will restrict ourselves to a neutral scalar field, and the extension can be straightforwardly done. Reviewing the existing literature, the study of the amplification of fundamental fields by rotating BHs can be traced back to Zel'Deovich\cite{Zel’Dovich1971, Zel’Dovich1972}, for a rotating cylindrical body in the non-GR premise. Subsequently, these analyses were developed further by Misner \cite{Misner1972}, Bekenstein\cite{Bekenstein:1973mi}, Teukolsky \cite{Teukolsky:1973ha, Teukolsky:1974yv} and Starobinsky \cite{Starobinskii:1973hgd, Starobinskil:1974nkd} to investigate the energy extraction process and stability of Kerr BH in response to external perturbations. More recent analyses on this topic can be found in \cite{Leite:2018mon, Leite:2017zyb, Brito:2015oca}.  Although there are various motivations to study superradiance (detailed discussion can be found in Ref.\cite{Brito:2015oca}), the amplification factor is particularly important and gained much interest in recent times as the greybody factor plays an important role in GW observables \cite{Oshita:2023cjz, Rosato:2024arw}. In the context of gravity theories coupled with NED, amplification factors for RBHs have been studied to explore the implications of the associated parameters and charge \cite{Yang:2022uze}. Similar studies for BHs emerging from various other modified gravity theories can be found in \cite{Liu:2024qso, Franzin:2022iai, Li:2022kch, Jha:2022ewi}. Therefore, our primary focus will be on evaluating the amplification factor for the massive scalar field, with the possibility that it could be generalized for the electromagnetic and gravitational fields in future. Moreover, the massive potential barrier for rotating spacetime also leads to superradiant instability \cite{Brito:2015oca}, hence, we have also explored the regime of the parameter space for GK BH leading to this instability.  

The remainder of the paper is organized as follows: Sec.\ref{GKBH} provides a brief overview of the GK BH and its characteristics. Next, Sec.\ref{sc.pert} presents the formulation of the minimally coupled scalar field perturbation equations and explores the separable radial and angular components of the equation of motion. In Sec.\ref{def.amp.sc}, we have discussed the asymptotic nature of the perturbation solution and have defined the amplification factor. This section also serves to find out the condition for superradiant amplification of the scalar field. After that, in Sec.\ref{numerics}, we have outlined the methodology for numerically computing the amplification factor and presented the results along with an explanation of their characteristics.
In Sec.\ref{instable.GK}, the superradiant instability regime of the GK BH due to massive perturbation has been argued following an analytical approach. Finally, we have concluded with potential future directions in Sec.\ref{concl}.

In the following analysis, we will adopt the metric signature $(-,+,+,+)$ and work in natural units, setting $\hbar = c = G = 1$.

\section{Magnetically charged rotating Ghosh-Kumar black hole}\label{GKBH}
In this section, we will briefly describe the magnetically charged BH proposed by Ghosh and Kumar \cite{Ghosh:2021clx}, known as GK BH \cite{Vagnozzi:2022moj}. The following action describes the standard GR theory coupled with nonlinear electrodynamics (NED) \cite{Fan:2016hvf},
\be
S=\int d^4x \sqrt{-g}\left[\frac{1}{16\pi}R-\frac{1}{4\pi}\mathcal{L}(\mathcal{F})\right],
\ee
where $R$ represents the Ricci scalar, and the NED Lagrangian denoted by $\mathcal{L}$, is a function of the electromagnetic field strength $\mathcal{F} = F_{\mu\nu}F^{\mu\nu}/4$. The explicit form of this Lagrangian, as considered by \cite{Ghosh:2021clx}, reads as, 
\be\label{mat.Lagrangian}
\mathcal{L}=\frac{4M\sqrt{q}\mathcal{F}^{5/4}}{q\left(\sqrt{2}+2q\sqrt{\mathcal{F}}\right)^{3/2}},
\ee
where $q$ is the magnetic charge and $M$ is the BH mass.
This Lagrangian, notably, behaves as $\mathcal{L}\sim \mathcal{F}^{5/4}$ in the weak field regime ($\mathcal{F}\to 0$), indicating a persistent non-linearity. Since the key motivation often lies in eliminating the curvature singularity in the strong field regime, consistency in the weak field regime is occasionally relaxed. For example, the first proposal by Bardeen \cite{Bardeen1968}, later brought on a physical footing \cite{Ayon-Beato:2000mjt} as being sourced by magnetic monopole, compromised this aspect. Importantly, {\it in the weak-field regime}, the asymptotic behaviour of the Bardeen solution deviates from reproducing the Coulomb charge and the resulting potential has been interpreted as originating from a monopole charge associated with the self-gravitating magnetic field in NED \cite{Ayon-Beato:2000mjt}.  It is worth mentioning that, this similar feature also appears in the case of Hayward BH solution \cite{Hayward:2005gi}. We will argue this issue further in the following discussion. For now, let us study the BH solution arising from the above Lagrangian. 

\begin{figure}
\includegraphics[width=\linewidth]{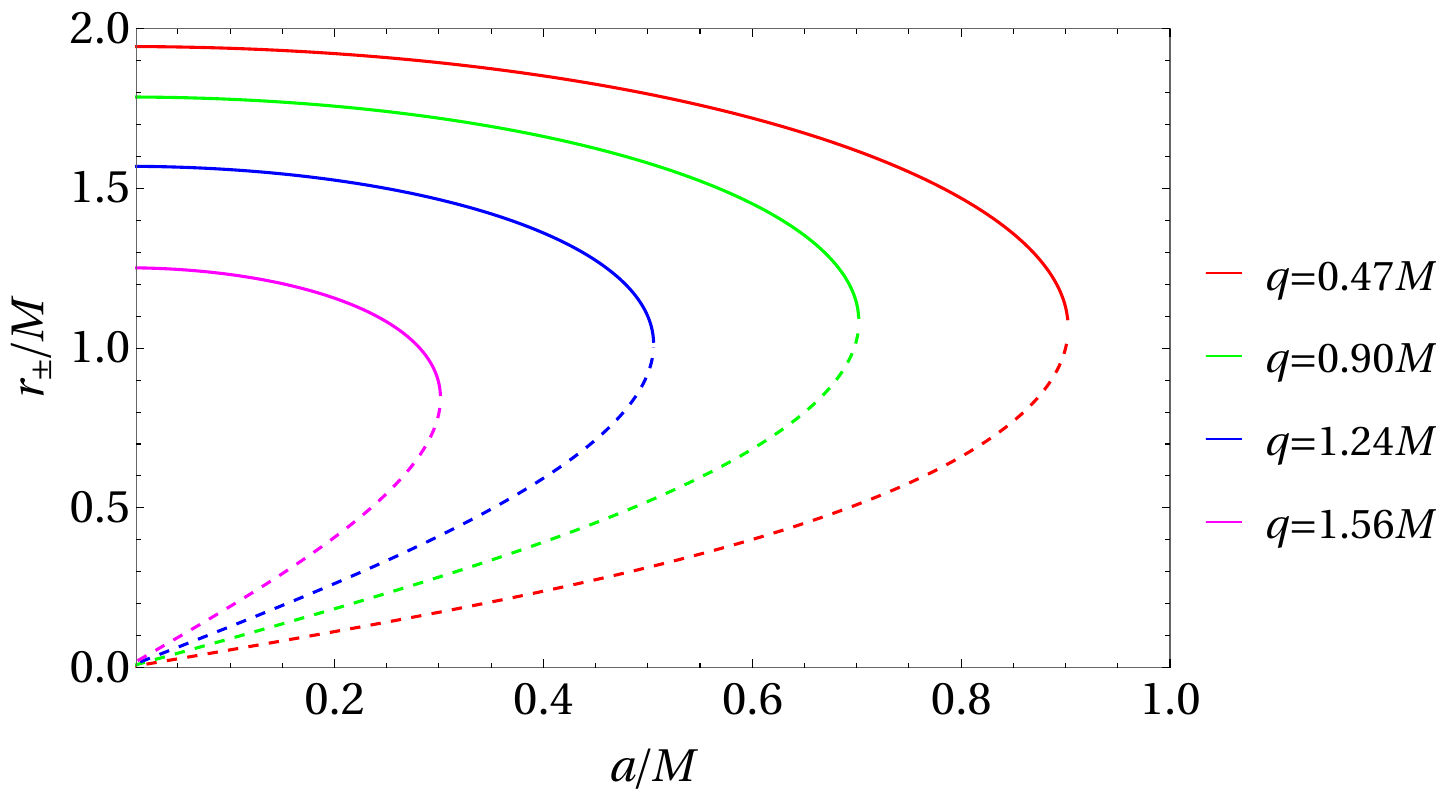}
\caption{The radius of the event horizon ($r_+$) represented by the solid lines and the Cauchy horizon ($r_-$) represented by the dashed line. Variation have been shown with the spin parameter, $a$, for the GK BH with different magnetic charge, $q$.}\label{horizon_plt}
\end{figure}

The method to derive the rotating solution begins with obtaining the static, spherically symmetric solution for the given Lagrangian. Then applying the modified version of Newman-Janis algorithm \cite{Newman:1965tw, Azreg-Ainou:2014pra}, one will be able to obtain the rotating version of the solution. In Boyer-Lindquist coordinate, the line element of rotating magnetically charged GK BH can be expressed as,
\be\label{GK.metric}
\bea
ds^2&=-\left(1-\frac{2M r^2}{\Sigma (r^2+q^2)^{1/2}}\right)dt^2+\frac{\Sigma}{\Delta}dr^2+\Sigma(r,\theta) d\theta^2\\
&~~~~~~~~~~~-\frac{4aMr^2\sin^2\theta}{\Sigma (r^2+q^2)^{1/2}}dtd\varphi+\frac{A(r,\theta)}{\Sigma}\sin^2\theta d\varphi^2,
\eea 
\ee
where 
\be
\bea
&\Sigma=r^2+a^2\cos^2\theta,\\
&\Delta=r^2-\frac{2Mr^2}{(r^2+q^2)^{1/2}}+a^2,\\ 
&A(r,\theta)=(r^2+a^2)^2-a^2\Delta \sin^2\theta.
\eea
\ee
Like Kerr BH \cite{Padmanabhan:2010zzb}, this spacetime also has ring shape curvature singularity for $\Sigma=0$ and coordinate singularity for $\Delta=0$. Therefore, the radius of the horizons can be determined in the usual manner by finding the root of $\Delta=0$. Due to the presence of the complicated form of the mass function, we have followed the numerical root-finding method, implemented in Mathematica. The behaviours of the event horizon ($r_+$) and the Cauchy horizon ($r_-$) have been illustrated in Fig.\ref{horizon_plt}, with the spin, $a$, for various values of the magnetic charge $q$. We have also presented the critical values of the parameters, spin $a$ and $q_c$ in Table-\ref{GK.critq}, for which the spacetime is allowed to have two horizons. Note that, in most of the following analysis, we will use the value of the parameters in such a way that $q\leq q_c$ is satisfied. For a more detailed analysis of the properties of GK BH, we refer the reader to \cite{Ghosh:2021clx}. 

Before proceeding, it is worth mentioning that on one hand the persisting non-linearity of the underlying theory \eqref{mat.Lagrangian} in the weak field, as also mentioned before, should be cured, still, investigations of the singular GK BH in the strong field regime can be accurately performed (see \cite{Macedo:2014uga} for Bardeen BH) to study the implications of the NED in the observables. However, one issue in doing so is that the observables are often measured far away from the BH, i.e., in the weak field regime, therefore inconsistency in the asymptotic regime could play a role, however, small. In our study, we primarily focus on superradiant amplification, whose underlying mechanism is based on the existence of the ergosphere in the strong field scenario \cite{Brito:2015oca}. Furthermore, due to the consideration of a neutral scalar field, we will show in the later sections that the persistent non-linearities in the weak field regime (relevant at spatial infinity) have no impact on the field modes as they asymptotically decouple from the background spacetime.
\renewcommand{\arraystretch}{1.5}
\begin{table}[h!]
    \centering
       \begin{tabular}{|wc{6.4em}|wc{4.3em}|wc{4.3em}|}
        \hline
        \multirow{2}{*}{Spin-paramter} & \multicolumn{2}{|c|}{Critical charge $(q_c/M)$}\\ 
        \cline{2-3} ($a/M$) &GK & KN\\
        \hline
        0.9  & 0.47 & 0.43\\ 
        \hline
        0.7 & 0.90 & 0.71\\ 
        \hline
        0.5 & 1.24 & 0.86\\ 
        \hline
        0.3 & 1.56 & 0.95\\ 
        \hline
    \end{tabular}
    \caption{Critical values of the magnetic charge parameter, $q$ for different values of spin of GK BH. We have also provided the corresponding values for the KN BH for comparison.}
    \label{GK.critq}
\end{table}
\section{Massive scalar test field in the background of the rotating Ghosh-Kumar black hole}\label{sc.pert}
We consider the minimally coupled action for the massive scalar field in the background of the GK BH spacetime, so that, the equation of motion can be written as,
\be
-\frac{1}{\sqrt{-g}}\pr_\mu\left(\sqrt{-g}g^{\mu\nu}\pr_\nu \Phi\right)+\mu^2\Phi=0,
\ee
where, $\mu$ in the last term symbolizes the mass of the scalar field. Now, the background metric \eqref{GK.metric} allows for two Killing vectors, $\xi_{(t)}=\pr_t$ and $\xi_{(\varphi)}=\pr_\varphi$, associated with $t$ and $\varphi$ respectively \cite{Padmanabhan:2010zzb}. Therefore, we decompose the scalar field in the following manner, \cite{Starobinskii:1973hgd},
\be
\Phi(t,r,\theta,\varphi)=e^{-\mi\omega{t}}e^{\mi{m}\varphi}R_{lm}(r)S_{lm}(\theta),
\ee
where $l$ and $m$ denote the orbital and azimuthal mode of the scalar field respectively. Whereas, $\omega$ represents the frequency of the scalar field. Substituting the above decomposition in the equation of motion, we get two separate equations, for $S_{lm}(\theta)$ and $R_{lm}(r)$. The equation governing $R_{lm}(r)$ turns out as,
\be\label{eq.rad}
\bea
&\Delta\frac{d}{dr}\left(\Delta\frac{dR_{lm}(r)}{dr}\right)+\Big[\left(\omega(r^2+a^2)-am\right)^2\\
&~~~~~~~~~~~-\Delta\left(\omega^2a^2+\mu^2r^2-2am\omega+\Lambda_{lm}\right)\Big]R_{lm}(r)=0,
\eea
\ee
with $\Lambda_{lm}$, the separation constant will be discussed momentarily. And the equation corresponding to $S_{lm}(\theta)$ can be expressed as,
\be\label{eq.sph}
\bea
&\frac{1}{\sin\theta}\frac{d}{d\theta}\left(\sin\theta\frac{d}{d\theta}{S_{lm}(\theta)}\right)+\Big\{a^2(\omega^2-\mu^2)\cos^2\theta\\
&~~~~~~~~~~~~~~~~~~~~~~~~~~~~~~~~~-\frac{m^2}{\sin^2\theta}+\Lambda_{lm}\Big\}S_{lm}(\theta)=0.
\eea
\ee
The physical solution of (\ref{eq.sph}) is well studied in literature\cite{Detweiler:1973zz}\cite{Brill:1972xj} and corresponds to spheroidal-eigenfunction. It resembles the usual Legendre's function in the limit $a\sqrt{\omega^2-\mu^2}\to{0}$. In the general case, the eigenvalue can be written as a power series expansion \cite{Seidel:1988ue, Berti:2005gp},
\be
\Lambda_{lm}=l(l+1)+\sum_{n=1}^\infty f_n \left(a\sqrt{\omega^2-\mu^2}\right)^n,
\ee
where $f_n$ are constant coefficients, which depend on the mode of the scalar field $(l,m)$. Up to $n=6$, the expression of the coefficients, $f_n$, have been given in Ref.\cite{Seidel:1988ue}. For our numerical analysis, we have considered up to the same order ($n=6$) with the restriction that $a\sqrt{\omega^2-\mu^2}<1$. Moreover, we find for the scalar field, $f_n=0$ with odd $n$. Also, note that we do not need to evaluate the spheroidal harmonics as the eigenvalues will be enough to evaluate the amplification factor. Next, we will define this factor by studying the asymptotic nature of the radial solution.
\section{Definition of the amplification factor for scalar field}\label{def.amp.sc}
The radial equation (\ref{eq.rad}) can simply be solved in the near horizon and near infinity limit by parametrizing the variable in the following way \cite{Starobinskii:1973hgd},
\be
\frac{dr_*}{dr}=\frac{r^2+a^2}{\Delta} ~,~ R(r)=\frac{u(r)}{\sqrt{r^2+a^2}},
\ee
where $r_*$ represents the tortoise coordinate. With this setup, the radial equation \ref{eq.rad} takes the following form, which resembles Schr\"odinger-like wave equation,  
\be\label{u.eq}
\frac{d^2 u(r_*)}{dr^2_*}+V_{\rm eff}(r)u(r_*)=0,
\ee
where 
\be
\bea
&V_{\rm eff}(r)=\Big(\omega-\frac{am}{r^2+a^2}\Big)^2-\frac{\Delta}{(a^2+r^2)^2}\Big[(a^2\omega^2+\mu^2r^2\\
&-2am\omega+\Lambda_{lm})+\sqrt{r^2+a^2}\frac{d}{dr}\left\{\frac{\Delta r}{(r^2+a^2)^{3/2}}\right\}\Big].
\eea
\ee
In the asymptotic limits ($r_*\to \pm \infty$), this effective potential read as,
\be
\bea
&{
V_{\rm eff}(r)= \left\{ 
        \begin{array}{l} 
           ~~~~~~ \omega^2-\mu^2,\hspace{1cm}r_*\to{\infty} (r\to\infty),\\
            \left(\omega-m\Omega_+\right)^2,\hspace{1cm}r_*\to{-\infty} (r\to{r_+}),\\
        \end{array}
        \right.}
\eea
\ee
where $\Omega_+=a/(r^2_++a^2)$ represents the horizon angular velocity. Substituting the above form of the potential in \eqref{u.eq}, one will be able to obtain the asymptotic solutions. To study the amplification process by BH spacetime, one needs to impose the boundary conditions, such that the scalar wave is purely ingoing near the horizon and superposition of ingoing and outgoing near special infinity \cite{Starobinskii:1973hgd}. With this in mind, the asymptotic solutions can be expressed in the following manner,
\be\label{u.bc}
{u(r)= \left\{ 
        \begin{array}{l} 
           \mathcal{I}_{\omega lm} e^{-\mi k_\infty {r_*}}+\mathcal{R}_{\omega lm}e^{\mi k_\infty {r_*}},~~r_*\to{\infty} (r\to\infty)\\
          ~~~\mathcal{T}_{\omega lm}e^{-\mi k_+ r_*},~~~~~~~~~~~~~~~r_*\to -\infty (r\to r_+),\\
        \end{array}
        \right.}
\ee
where $k_\infty=\sqrt{\omega^2-\mu^2}$ and $k_+=(\omega-m\Omega_+)$. Whereas, $\mathcal{I}_{\omega lm}, \mathcal{R}_{\omega lm}$ and $\mathcal{T}_{\omega lm}$ symbolize the incidence, reflection and  transmission coefficients respectively. Importantly, one can see that the field modes in the spatial infinity decouple from the background spacetime as mentioned earlier. Therefore the persisting non-linearity of the underlying NED theory in the weak field regime does not have much impact. However, the effect of NED only enters through the strong field regime (near horizon dynamics) and thereby affects the coefficients of field modes in the spatial infinity. Hence the observables will encode the feature of NED through these coefficients. Now, from the above equation, it can be realized that the condition $\omega>\mu$ should be maintained for scattering states. The Wronskian, $W[u,u^*]=u\pr_{r_*}u^{*}-u^*\pr_{r_*}u$, of \eqref{u.eq} turns out to be
\be
W[u,u^*]= {\left\{ 
        \begin{array}{l} 
           2\mi\sqrt{\omega^2-\mu^2}(|\mathcal{I}_{\omega lm}|^2-|\mathcal{R}_{\omega lm} |^2),~~r_*\to \infty\\
          ~~~2\mi(\omega-m\Omega_+)|\mathcal{T}_{\omega lm} |^2, ~~r_*\to -\infty.\\
        \end{array}
        \right.}
\ee
Then the conservation of the Wronskian leads to the following condition based on the above equation, 
\be
(|\mathcal{I}_{\omega lm}|^2-|\mathcal{R}_{\omega lm}|^2)=\frac{(\omega-m\Omega_+)}{\sqrt{\omega^2-\mu^2}}|\mathcal{T}_{\omega lm} |^2.
\ee
The above formula implies the following condition for superradiant amplification \cite{Brito:2015oca},
\be\label{amp.cond}
\mu<\omega<m\Omega_+
\ee
This implies that the scalar field will be amplified due to the rotation in the spacetime, $a\neq 0$, otherwise $\Omega_+$ will be zero. However, due to the presence of magnetic charge in GK BH, the event horizon will be modified and in turn, will affect the horizon angular velocity, $\Omega_+$. The maximum frequency, $\omega=m\Omega_+$, at which superradiance occurs is often referred to as the \textit{cutoff frequency} for superradiance. Nevertheless, we quantify the amplification factor from the above equation as \cite{Brito:2015oca},
\be\label{def.amp}
Z_{lm}(\omega)=\left|\frac{\mathcal{R}_{\omega lm}}{\mathcal{I}_{\omega lm}}\right|^2-1.
\ee
Next, we will discuss the evaluation process of the above expression of amplification factor and analyse the effect of the magnetic charge.
\begin{figure}[t]
\includegraphics[width=\linewidth]{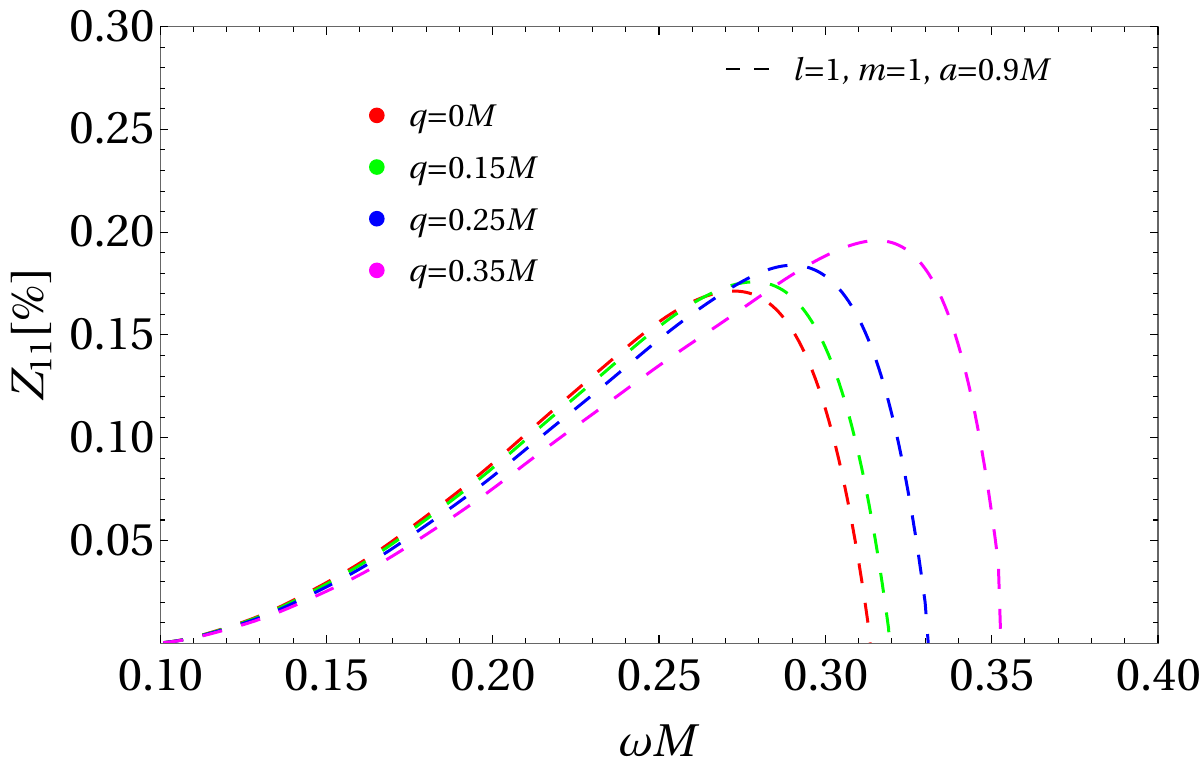}
\caption{The amplification factor of the scalar field with frequency $\omega$ has been plotted for the superradiant mode, $l=1$ and $m=1$, considering various charge values $q$ for the GK BH at a fixed spin. One can notice that the maximum amplification is $\sim 0.2$\%. 
Notably, we have used the compactified notation in the labelling of this plot, as well as in other figures, for the betterment of the representation as done in \cite{Li:2022kch}, and elsewhere. To get an idea of the typical magnitude of superradiant amplification, one may take a look at Ref. \cite{Franzin:2022iai}, where actual values have been used for the labelling.}\label{GK_0p9_amp}
\end{figure}
\begin{figure*}[t]
\includegraphics[width=\linewidth]{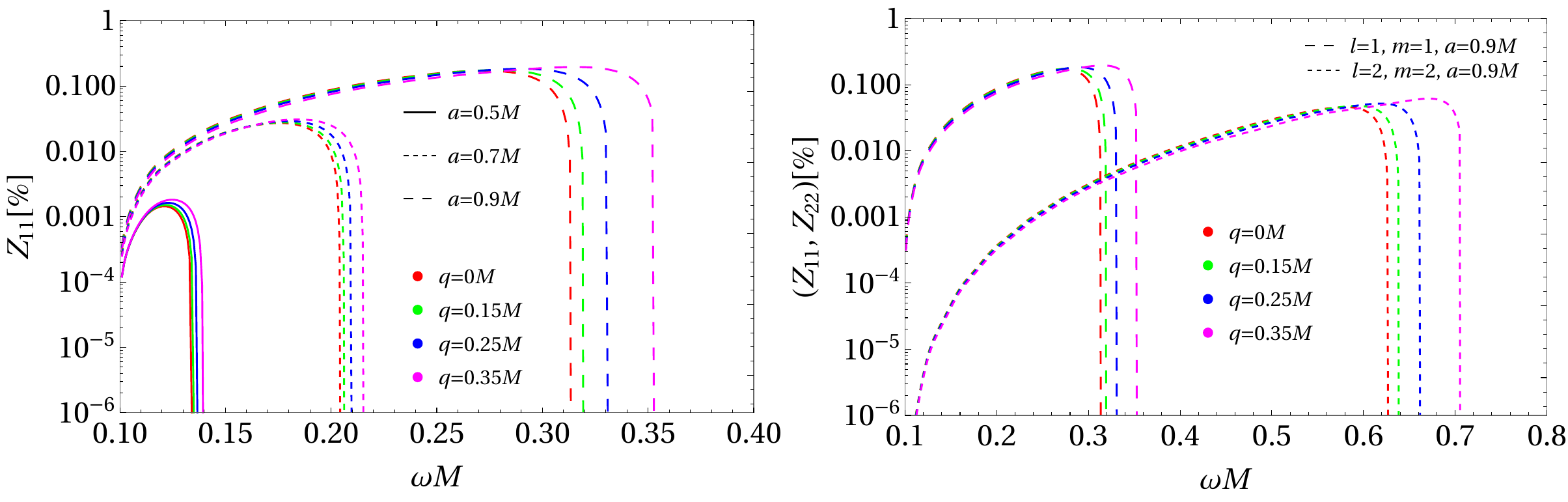}
\caption{Amplification factor of the scalar field with frequency, $\omega$: In the left panel, we have shown the results for the dominant superradiant mode of the scalar field, $l=1, m=1$, by considering different values of the spin parameter, $a$, with a common set of allowed values of charge, $q$ for GK BH. In the right panel, we have compared the results for the two consecutive superradiant modes of the scalar field as described in the plot, by considering the same set of allowed charges, $q$ with a fixed value of the spin parameter, $a=0.9 M$ for GK BH.}\label{GK_amp}
\end{figure*}
\begin{figure*}[t]
\includegraphics[width=\linewidth]{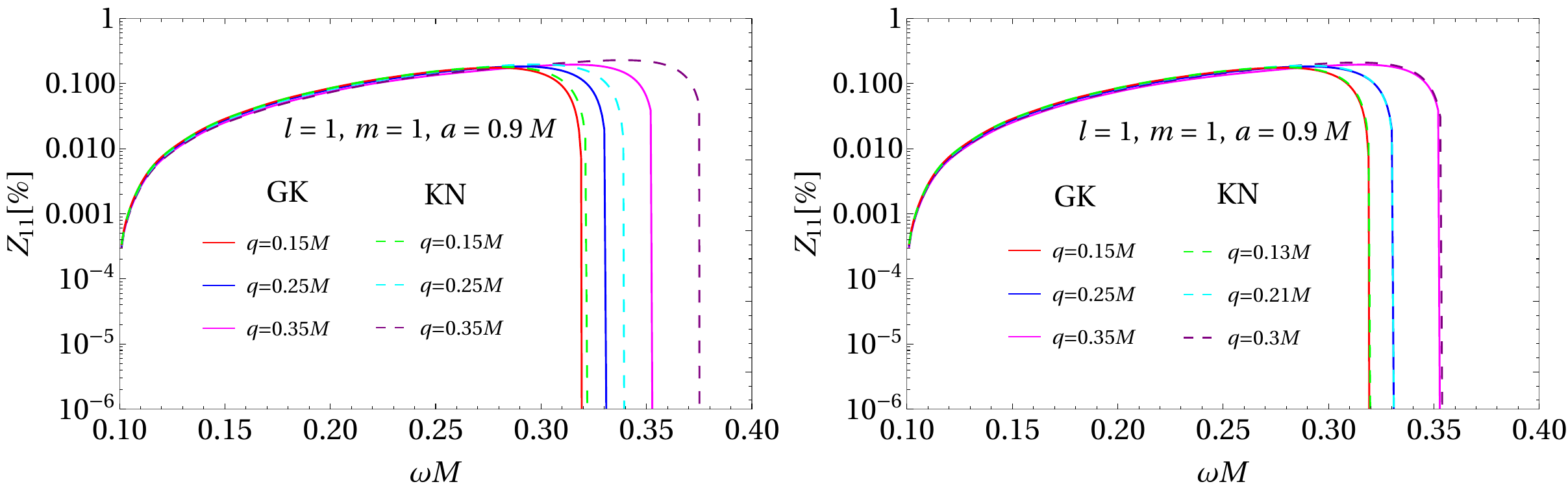}
\caption{Amplification factor of the scalar field with frequency, $\omega$ and dominant superradiant mode, $l=1, m=1$: In the left panel, we have presented the results for GK and KN BH by considering a common set of allowed charges, $q$ for a fixed value of the spin parameter, $a=0.9 M$. In the right panel, we have attempted to match the results for the two BHs with a fixed value of the spin parameter, $a=0.9 M$, by considering different set of allowed charges, $q$. Important to note that $q$ represents the electric charge for the KN BH, while it denotes the magnetic charge for the GK BH.}\label{GK_KN_amp}
\end{figure*}
\section{Numerical methodology and results}\label{numerics}
In this section, we will outline the numerical procedure used to evaluate the amplification factor for the massive scalar field in GK BH spacetime (for existing literature in this regard see \cite{Brito:2015oca}, and references therein). First, we numerically solve the perturbation equation \eqref{u.eq} with the ingoing initial condition \eqref{u.bc} set close to the event horizon at $r\to r_+(1+\epsilon)$ with $\epsilon<<1$. Next, we obtain the solution at $r_\infty=150 M$, which is located far from the event horizon. However, changing this evaluation point up to a certain extent does not affect the stability of our final results. Once we find out the numerical solution at the point approximating spatial infinity, we match it with the asymptotic form obtained for $r\to \infty$ \eqref{u.bc} and determine the reflection and incident coefficients. Next, by substituting these coefficients for different modes ($l, m$) into the definition \eqref{def.amp}, the mode-wise amplification factor for the scalar field can be determined with frequency, $\omega$. Important to note that there are three main theory parameters involved in the numerical analysis,: mass of the scalar field, $\mu$, magnetic charge, $q$ and spin of the BH, $a$; other than the modes of the scalar field ($\omega, l, m$). If otherwise not specified we have kept the mass of the scalar field fixed, $\mu=0.1/M$ in the following discussion. 

In Fig.\ref{GK_0p9_amp}, we have illustrated the amplification factor of the massive scalar field as a function of frequency $\omega$ for the superradiant mode $l=1, m=1$ with different values of magnetic charge, $q$ and a fixed value of the spin parameter, $a=0.9M$. We observe that with the increase of the magnetic charge $q$ of GK BH, the magnitude of amplification decreases, although not significantly, in a wide range of frequencies. However, near the cutoff frequency, the magnitude slightly increases with charge, before approaching zero. This feature also has been found before for rotating RBHs \cite{Yang:2022uze}. Another important feature is that increasing the magnetic charge significantly widens the cutoff frequency for superradiant amplification as compared to Kerr BH (which corresponds to the $q=0$ case, as shown in the plot). This can be understood from the fact that the radius of the event horizon for GK BH is lower than the Kerr BH, therefore the cutoff frequency, $m\Omega_+$ is higher. This is a typical feature of such charged BHs, as has been found out before \cite{Leite:2017hkm, Yang:2022uze, Benone:2019all}.

To see the behaviour for different values of the spin of GK BH, we have demonstrated in the left panel of Fig.\ref{GK_amp} the amplification factor of the scalar field for various values of the spin parameter with a common set of magnetic charges. As expected, higher spin values will further amplify the scattered scalar field and influence the cutoff frequency for superradiance via the horizon-angular velocity $\Omega_+$. To compare the characteristics of amplification for different modes, we have presented two consecutive superradiant modes, $l=1,m=1$ and $l=2,m=2$ in the right panel of Fig.\ref{GK_amp}. Note that the associated values of $m$ correspond to the maximum amplification for given $l$. For both plots, we have considered a common set of charges for a fixed spin parameter, $a=0.9 M$, as has been displayed in the figure. We see that the $l=1,m=1$ is the dominant superradiant mode as far as the magnitude of amplification is concerned. 

Having found various features of GK BH mimicking that of the KN BH, we have been motivated to explore the potential resemblance between these two BHs. First, we analyzed the deviations in the amplification factors between the GK and KN BHs using a common set of charges for a fixed spin parameter. The differences have been illustrated in the left panel of Fig. \ref{GK_KN_amp}.  It is important to note that $q$ represents the electric charge for the KN BH, while it denotes the magnetic charge for the GK BH. For the same charge, the radius of the event horizon for GK BH is higher than that of KN BH. Therefore, the allowed frequency range, obtained in \eqref{amp.cond}, is comparatively lower for GK BH. Next, to match the results, we find it suitable to decrease the value of charges for the KN BH. The final results have been shown in the right panel of Fig.\ref{GK_KN_amp}. We see that for low to moderate values of the charges, GK BH can potentially mimic the KN BH. It is worth mentioning that this is not the first time a correspondence has been studied between BHs in linear electrodynamics and those emerging from NED \cite{Macedo:2014uga, dePaula:2023muc, Karmakar:2024hng}. However, while these studies have focused on static BH spacetimes, we have explored this scenario in a rotating case for the first time.
\section{Analysis of superradiant instability regime for Ghosh-Kumar black hole}\label{instable.GK}
\begin{figure}[t]
\includegraphics[width=\linewidth]{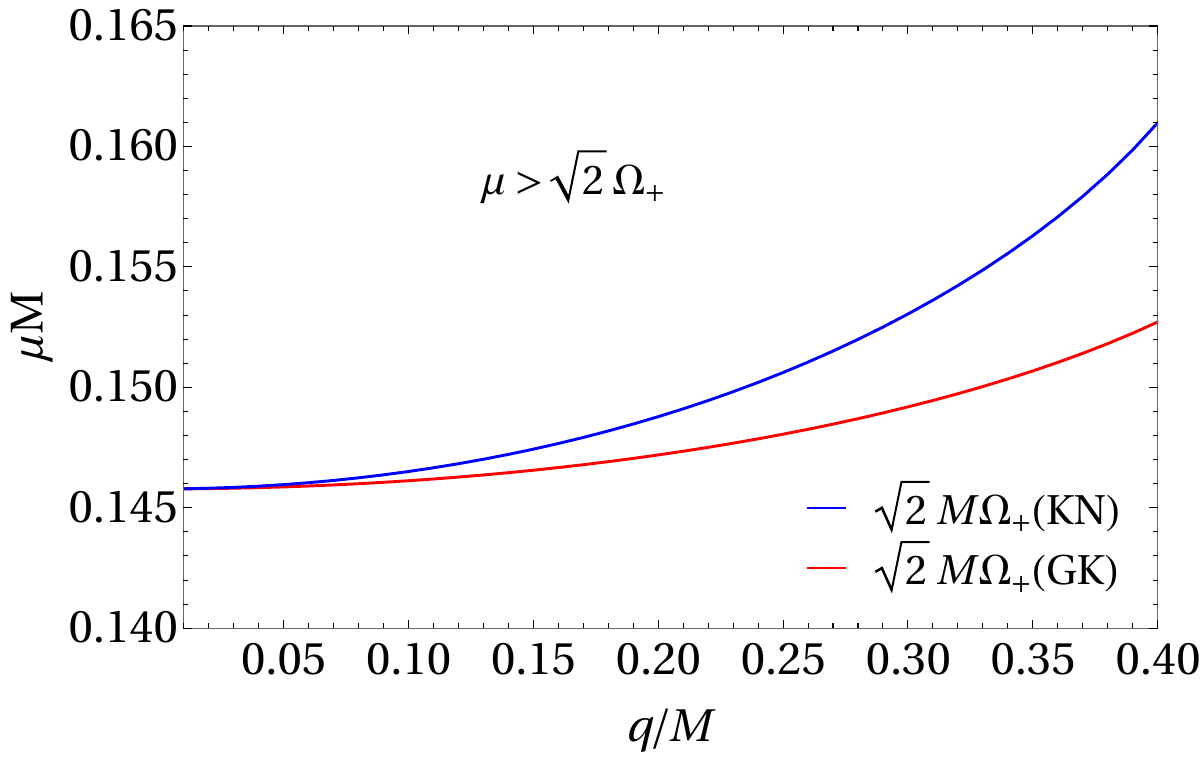}
\caption{The red and blue solid lines respectively represent the plot of $\sqrt{2} M\Omega_+$ for GK and KN BH. In the $Y$-axis, the labelling with the mass of the scalar field $\mu$ signifies the regime of instability depending on its value greater or less than $\sqrt{2}\Omega_+$, above and below the solid lines respectively, for a specific charge. We have particularly considered the dominant superradiant mode, $l=m=1$, of the scalar field for this plot.}\label{instabl.space}
\end{figure}
In the preceding sections, we have seen that the scattered scalar wave is being amplified as far as $\mu<\omega<m\Omega$, and analyse the impact of the magnetic charge of GK BH. The mass of the scalar field thus far attributed a restriction on the frequency in determining the scattering states. However, the presence of the mass term in the effective potential has an important implication as it leads to the formation of bound states and acts as a reflecting barrier alongside the angular momentum barrier \cite{Cardoso:2004nk}. An amplified scalar wave will be reflected by this mass barrier, and through repeated amplification this process can potentially result in dynamical instability, which is commonly known as the BH bomb \cite{Press:1972zz}. In the following section, we will explore the regime of parameter space for this instability. First, we rewrite the equation \eqref{eq.rad} in the following manner,
\be\label{redef.Reqn}
\bea
&\Delta\frac{d}{dr}\left(\Delta\frac{dR_{lm}(r)}{dr}\right)+\alpha(r)R_{lm}(r)=0,
\eea
\ee
with
\be
\bea
&\alpha(r)=\left(\omega(r^2+a^2)-am\right)^2\\
&~~~~~~~~~~~~~~~~-\Delta\left(\omega^2a^2+\mu^2r^2-2am\omega+\Lambda_{lm}\right).
\eea
\ee
The boundary conditions for bound state solutions can be expressed as \cite{Cardoso:2004nk},
\be
{R_{lm}(r)= \left\{ 
        \begin{array}{l} 
            \frac{e^{-\left(\sqrt{\mu^2-\omega^2}\right)r_*}}{r},~~~~~~~r_*\to{\infty} (r\to\infty)\\
          ~~~e^{-\mi(\omega-m\Omega_+)r_*},~~~~~~~~r_*\to -\infty (r\to r_+),\\
        \end{array}
        \right.}
\ee
with $\omega< \mu$. To analyze the instability, it will be helpful to redefine the radial part of the scalar field as, 
\be
R_{lm}(r)=\frac{\xi_{lm}(r)}{\sqrt{\Delta(r)}}.
\ee
Substituting the above form we obtain from \eqref{redef.Reqn} the governing equation of the new function $\xi_{lm}$, 
\be
\frac{d^2\xi_{lm}}{dr^2}+\left[\omega^2-V_\xi\right]\xi_{lm}=0,
\ee
where, 
\be
\omega^2-V_\xi=\frac{\alpha(r)+\beta(r)}{\Delta^2},
\ee
with
\be
\beta(r)=\left[-\frac{\Delta(r)\Delta''(r)}{2}+\frac{(\Delta'(r))^2}{4}\right].
\ee
Here, prime denotes the derivative with respect to $r$. For instability, the massive modes should be trapped well outside the event horizon that implies the effective potential should be increasingly flat as one approaches spatial infinity, $V'_\xi(r)\to 0^+$ as $r_*\to \infty$ \cite{Hod:2012zza}. To realize that, we first expand the potential in the asymptotic limit, which can be expressed as,
\be
V_\xi=\mu^2+\frac{\mu^2-2\omega^2}{r}+\mathcal{O}(1/r^2).
\ee
Then the condition for the trapping potential can be quantified by the following formula,
\be
V'_\xi=\frac{2\omega^2-\mu^2}{r^2}>0.
\ee
Based on this inequality, it can be inferred that instability will occur when $\mu <\sqrt{2}\omega$. Hence, the bound state frequency for which instability may happen is restricted as follows,
\be
\frac{\mu}{\sqrt{2}}<\omega<\mu.
\ee
Along with the condition of amplification, previously derived, the criteria for superradiant instability finally turns out as \cite{Hod:2012zza},
\be
\frac{\mu}{\sqrt{2}}<\omega<m\Omega_+.
\ee
From the above inequality, one can realize that as far as $\mu >\sqrt{2}m\Omega_+$, the BH spacetime will be stable. Furthermore, we see that the magnetically charged GK BH does not explicitly modify the condition for superradiant instability, at least up to $\mathcal{O}(1/r^2)$ approximation. However, for a given value of spin and charge, the difference in the event horizon radius between the GK and other BH spacetime, such as KN BH will significantly affect the superradiant instability regime through the angular velocity $\Omega_+$. We have illustrated in Fig.\ref{instabl.space} the stability and instability region of GK BH along with the KN BH. 
\section{Conclusion and future outlook}\label{concl}
We have investigated the superradiant amplification of the scalar field for a magnetically charged rotating GK BH, focusing on the influence of the charge parameter on the amplification process. To carry out the computation, we followed a numerical approach well-established in the literature \cite{Brito:2015oca}. In summary, our results indicate that the magnetic charge significantly widens the frequency range leading to superradiant amplification as compared to the neutral Kerr BH. However, the magnitude of the amplification factor does not increase substantially in comparison. Nevertheless, the characteristics displayed by the neutral scalar field for the GK BH are quite similar to those of the KN BHs \cite{Leite:2017hkm}. This has motivated us to explore the parameter space for which there is a direct resemblance between the two BHs. In particular, we have observed that for a specific value of the spin parameter, the amplification factor for the KN BH matches that of the GK BH when the charge of the former is slightly lower than that of the latter. This may be understood from the fact that the radius of the event horizon is larger in the case of GK BH for the same charge, naturally, the cutoff frequency turns out to be lower than the KN BH. However, the difference between the charges leading to the correspondence of the GK and KN BH diminishes as we decrease the value of the charge.

Investigating the GK BH, a singular solution characterized by features of the NED, holds considerable significance, on top of the existing analyses of regular solutions in a similar context \cite{Yang:2022uze, Macedo:2014uga}. Observationally distinguishing the regular and singular black hole solutions may provide a pathway to exploring the existence of singularity.  To understand how the underlying features of the BH manifest in the observables, we resorted to considering the scalar field, which adapts to the symmetries and characteristics of the background spacetime much like other fundamental fields. Importantly, the scalar field often acts as a proxy field indicating possible scenarios for charged scalar, EM and GWs, and others (bosonic fields for the superradiance). Our analysis reveals that distinguishable features are already present in the superradiantly enhanced outgoing modes of the scalar field for singular BHs, making it crucial for conducting rigorous investigations to distinguish them from non-singular NED BHs. It also prompts that one immediate extension of our work should be considering the charged scalar field that could shed more light on the non-trivial interplay with the magnetic charge of the GK BH.

Observing the amplified scalar wave signal directly is challenging. Instead, the scalar field could be treated as an axion-like field and directly coupled with photons, resulting in the rotation of the plane of polarization \cite{Fedderke:2019ajk}. Through this rotation angle one can further investigate the possibility of detecting the features of GK BH encoded in the superradianlty enhanced outgoing axion following Ref.\cite{Fujita:2018zaj} and others. Of course, considering the coupling of photons directly with the BH background would be a much more robust approach for the detectability of the outgoing signal, such as, in the radio observations. Similarly, the case of GW can be examined, and the prospects for detection in current and future observations can be explored. However, for non-vacuum solutions, such as those involving NED, the EM waves and GWs do not lead to the Teukolsky-like differential equations \cite{Teukolsky:1973ha, Teukolsky:1974yv}. Therefore, a non-trivial numerical approach would be essential for performing the computation. We will address some of these issues in our future project.

Finally, we have analyzed the superradiant instability of the scalar field. From the analytical calculation, we find that the generic condition for this instability remains intact for the GK BH. However, the presence of the magnetic charge shifts the event horizon's radius in a different manner as compared to KN BH, which in turn affects the parameter space that leads to this instability. Nevertheless, the detailed analysis of the corresponding bound state frequencies should be investigated further.
We should also emphasize that the region of superradiantly amplified scalar-bound states is known for forming a cloud or condensate of particles \cite{Brito:2015oca}, which serves as a smoking gun for particle physics phenomenology \cite{Cardoso:2018tly, Baryakhtar:2017ngi, Baryakhtar:2020gao}. Significant effort is also being made to uncover the imprint of the strong gravity regime in the secondary continuous GWs emitted by this superradiant cloud \cite{East:2018glu}, both in current LVK \cite{LIGOScientific:2021rnv, Arvanitaki:2016qwi, Baumann:2018vus} and future GW observations \cite{Ghosh:2018gaw, Hannuksela:2018izj}. In principle, these observables should encode the signature of the key features of the underlying BH spacetime. Therefore, the current analysis presents a promising opportunity to explore these directions in the context of the GK BH.

\noindent
\textbf{Acknowledgments:}
The author would like to acknowledge the weekly discussion in the group meeting supervised by Debaprasad Maity. We also thank the anonymous referee for improving the manuscript by providing insightful comments and suggestions.

\end{document}